\documentclass[usenatbib,useAMS,usegraphicx]{mn2e}
\usepackage{txfonts}

\renewcommand\pi\upi

\newcommand\rhoDM{\rho_{\rm D}}
\newcommand\rhoL{\rho_\ell}
\newcommand\SigmaL{\Sigma_\ell}
\newcommand\Rd{R_{\rm d}}
\newcommand\rd{r_{\rm d}}
\newcommand\sigmalos{\sigma_\mathrm{los}}
\newcommand\bnabla{\bmath\nabla}
\newcommand\lap{\nabla^2}

\title[Cores and Cusps]
{Cores and Cusps in the Dwarf Spheroidals}
\author[Evans, An, \& Walker]
{N.~W.~Evans,$^1$\thanks{E-mail:
nwe, walker@ast.cam.ac.uk (NWE, MGW),
jinan@nbi.dk (JA)}
J.~An,$^{2,3}$\footnotemark[1] and
M.~G.~Walker$^1$\footnotemark[1]\\
$^1$ Institute of Astronomy, University of Cambridge, 
Madingley Road, Cambridge, CB3 0HA\\ 
$^2$ Dark Cosmology Centre,
Niels Bohr Institutet, K{\o}benhavns Universitet,
Juliane Maries Vej 30, DK-2100 Copenhagen~\O, Denmark\\
$^3$ Niels Bohr International Academy,
Niels Bohr Institutet, K{\o}benhavns Universitet,
Blegdamsvej 17, DK-2100 Copenhagen~\O, Denmark}

\date{to be submitted}
\pagerange{\pageref{firstpage}--\pageref{lastpage}}\pubyear{2008}

\begin{document}
\maketitle

\label{firstpage}


\begin{abstract}
  We consider the problem of determining the structure of the dark
  halo of nearby dwarf spheroidal galaxies (dSphs) from the spherical
  Jeans equations.  Whether the dark halos are cusped or cored at the
  centre is an important strategic problem in modern astronomy. The
  observational data comprise the line-of-sight velocity dispersion of
  a luminous tracer population. We show that when such data are
  analysed to find the dark matter density with the spherical Poisson
  and Jeans equations, then the generic solution is a dark halo
  density that is cusped like an isothermal ($\rhoDM \propto
  r^{-2}$). Although milder cusps (like the Navarro-Frenk-White
  $\rhoDM \propto r^{-1}$) and even cores are possible, they are not
  generic. Such solutions exist only if the anisotropy parameter
  $\beta$ and the logarithmic slope of the stellar density
  $\gamma_\ell$ satisfy the constraint $\gamma_\ell = 2\beta$ at the
  centre or if the radial velocity dispersion falls to zero at the
  centre. This surprisingly strong statement is really a consequence
  of the assumption of spherical symmetry, and the consequent
  coordinate singularity at the origin. So, for example, a dSph with
  an exponential light profile can exist in Navarro-Frenk-White halo
  and have a flat velocity dispersion, but anisotropy in general
  drives the dark halo solution to an isothermal cusp.  The identified
  cusp or core is therefore a consequence of the assumptions
  (particularly of spherical symmetry and isotropy), and not the data.
\end{abstract}



\section{Introduction}

Hot stellar systems are held up by the stellar velocity dispersion and
have little or no rotation. In fact, many such stellar systems --
giant elliptical galaxies and dwarf spheroidals -- have a velocity
dispersion profile that is constant to a good approximation. The case
of the dwarf spheroidals (dSphs) has received particular attention in
recent years. \citet{Kl01,Kl02} showed that the velocity dispersion
profile of Draco is flattish through out the bulk of the galaxy,
although later work by \citet{Wi04} found evidence for kinematically
cold populations in the very outermost parts.  \citet{Wa07} presented
stellar velocity dispersion profiles for seven Milky Way dSphs and
found almost all to be constant to a good approximation right in to
the very centre, although the profile of Sextans seems to dip
somewhat. \citet{Ko07a,Ko07b} studied Leo~I and Leo~II, and also found
essentially flattish profiles.

The hope has been that gathering line-of-sight velocities of bright giant
stars in the Milky Way dSphs may provide evidence of the structure of
the halo in these extremely dark matter dominated galaxies. A question
of great interest is whether the dark halos of the dSphs are cusped,
as predicted by numerical simulations in cold dark matter cosmogonies,
or cored -- but results so far have been inconclusive. For example,
\citet{Ko07b} found that the velocity dispersion data on Leo~II are
consistent with halo dark matter densities that are both cored and
cusped.  \citet{Wi06} and \citet{Gi07} presented an analysis of the
velocity dispersion data for six dSphs based on the Jeans equations
and argued that cored density profiles were favored, partly because
this also explains the persistence of kinematically cold substructure
in the Ursa Minor dSph \citep{Kl03} and the maintenance of globular
clusters in the Fornax dSph \citep{Go06}.

Here, we are less concerned with modelling observational data on any
given dSph than with understanding the generic qualities of the light
and dark matter profiles.  We consider the general problem of a tracer
stellar population with a known line-of-sight velocity dispersion
residing in a spherical dark matter halo of an unknown density
law. Given the data, Sect.~\ref{sec:jd} considers what can be
legitimately inferred concerning the properties of the dark
halo. Motivated by the flatness of the observed dispersion profiles,
Sect.~\ref{sec:cvd} considers constant velocity dispersion solutions
of the Jeans equations.  We show that the generic solution gives a
halo density law with an isothermal cusp ($\rhoDM \propto
r^{-2}$). Although other solutions are possible -- in particular with
cores or with the milder cusps preferred by cosmologists ($\rhoDM
\propto r^{-1}$) -- they are not generic. Sect.~\ref{sec:ex} gives
some examples, which show why Jeans modelling in the isotropic case
has yielded results consistent with both cores and cusps. In our
examples, however, any anisotropy drives the dark halo solution to the
isothermal cusp. Finally, in Sect.~\ref{sec:gen}, we discard the
assumption of constancy of the velocity dispersion. Solely within the
framework of the spherical symmetric Poisson and Jeans equations, we
show that solutions of these equations almost always possess
isothermal cusps, unless some very special conditions are satisfied
either by the radial velocity dispersion or by the anisotropy and the
logarithmic gradient of the light profile.

\section{The Jeans Degeneracy}
\label{sec:jd}

The observables are the surface brightness and the line-of-sight
velocity dispersion of a stellar population.  Given a mass-to-light
ratio ($\Upsilon$), the surface mass density of the stellar
populations ($\SigmaL)$ can be deduced from the surface brightness.
If the system is spherically symmetric, $\SigmaL(R)$ is then related
to the three-dimensional density associated with the luminous material
$\rhoL(r)$ via an Abel transform. Here $R$ is the projected distance,
whilst $r$ is the three-dimensional distance measured from the centre
of the halo.  The inverse transform provides us with the unique
$\rhoL$ \citep{BT};
\begin{equation}
\label{eq:integdeb}
\rhoL(r)=
-\frac1\pi\int_r^\infty\frac{\mathrm d\SigmaL}{\mathrm dR}
\frac{\mathrm dR}{\sqrt{R^2-r^2}}.
\end{equation}
However, even if one assumes spherical symmetry, the behaviour of the
line-of-sight velocity dispersion does not produce a unique solution
for the radial dependence of the radial and tangential velocity
dispersions.  The ``luminosity weighted'' (assuming a constant
$\Upsilon$ for the stellar population) line-of-sight velocity
dispersion $\sigmalos^2$ is given by the integral
\begin{equation}
\label{eq:losvd}
\SigmaL(R)\sigmalos^2(R)
=2\int_R^\infty\left(1-\beta\frac{R^2}{r^2}\right)
\frac{\rhoL\sigma_r^2r\,\mathrm dr}{\sqrt{r^2-R^2}}
\end{equation}
where $\beta=1-(\sigma_\theta^2/\sigma_r^2)$ is the anisotropy
parameter for a spherical system, and $\sigma_r$ and $\sigma_\theta$
are the radial and (one-dimensional) tangential velocity dispersions.
It has long been known \citep[see e.g.,][]{MK90,DM92} that the
line-of-sight velocity second moment is degenerate, in that there
exist many sets of solutions -- $\sigma_r^2(r)$ and $\beta(r)$ -- that
reproduce the observables.

For example, suppose that we have $\rhoL(r)$ from
equation (\ref{eq:integdeb}). Then, for any given behaviour
of $\sigma_r^2(r)$, the anisotropy parameter $\beta(r)$ can be
found\footnote{The first part of eq.~(\ref{eq:abelbeta})
is easily verified after re-arranging eq.~(\ref{eq:losvd})
and applying an inverse Abel transform. The derivation of
the second part requires switching the order of a double
integral and performing explicit integrations on it.}
to reproduce the observables such as
%
\begin{eqnarray}
\beta(r)&=&
-\frac1\pi\frac{r^2}{\rhoL\sigma_r^2}
\int_r^\infty\frac{\mathrm dR}{\sqrt{R^2-r^2}}\,
\frac{\mathrm d}{\mathrm dR}\!
\left[\frac{\SigmaL(\bar\sigma_r^2-\sigmalos^2)}{R^2}\right]
\nonumber\\&=&
1+\frac1{\rhoL\sigma_r^2r}
\int_r^\infty\!\mathrm d\tilde r\,\rhoL(\tilde r)\sigma_r^2(\tilde r)
\nonumber\\&&+\frac{r^2}{\pi\rhoL\sigma_r^2}
\int_r^\infty\frac{\mathrm dR}{\sqrt{R^2-r^2}}\,
\frac{\mathrm d}{\mathrm dR}\!
\left(\frac{\SigmaL\sigmalos^2}{R^2}\right)
\label{eq:abelbeta}
\end{eqnarray}
%
where
\[
\bar\sigma_r^2(R)
=\frac2{\SigmaL(R)}
\int_R^\infty\frac{\rhoL\sigma_r^2r\,\mathrm dr}{\sqrt{r^2-R^2}}.
\]
is the luminosity-weighted mean projected radial velocity
dispersion. [We have not found eq.~(\ref{eq:abelbeta}) for the
anisotropy parameter in the existing literature.] Not all solutions
are necessarily physical since $-\infty \le \beta \le 1$, which has to
be checked \emph{a posteriori}. However, subject to this condition,
\emph{for any given arbitrary $\sigma^2_r(r)$, an anisotropy parameter
  $\beta(r)$ can be found to reproduce any observable
  $\sigmalos^2(R)$.}

Once we have $\rhoL(r)$, guessed (any) $\sigma_r^2(r)$, and
found the anisotropy parameter $\beta(r)$ to reproduce the
observables, then from the spherically-symmetric steady-state
Jeans equation
\begin{equation}
\label{eq:ssj0}
\frac{\mathrm d(\rhoL\sigma_r^2)}{\mathrm dr}
+2\beta\frac{\rhoL\sigma_r^2}r
=-\frac{4\pi G\rhoL}{r^2}\!
\int_0^r\!\rho_{\rm t}(\tilde r)\,\tilde r^2\,\mathrm d\tilde r.
\end{equation}
%
%
%
Here, $\rho_{\rm t}$ is the total density such that
%
%
$\rho_{\rm t}=\Upsilon\rhoL+\rhoDM$
where $\rhoL$ and $\rhoDM$ are
the stellar and dark matter densities, respectively.
With the self-consistency assumption, if there is no dark matter
($\rhoDM=0$), and the mass-to-light ratio $\Upsilon$ is
additionally constant, then the choice of $\sigma_r^2(r)$ is
uniquely determined by the coupled Poisson and Jeans
equations \citep[see e.g.,][]{Bi82,To83,DM92}.

However, if $\Upsilon$ is varying or $\rhoDM(r)\ne 0$,
there is no unique choice of $\sigma_r^2(r)$.  That is, any
$\sigma_r^2(r)$ is allowed subject only to the constraints that
$-\infty \le \beta(r) \le 1$ and $\rho_{\rm t}(r)\ge0$.
Consequently, without further observational constraints or simplifying
assumptions, determining a model to reproduce the observed
$\SigmaL(R)$ and $\sigmalos^2(R)$ is completely
indeterminate in the spherical case.


\section{Jeans Modelling with Constant Velocity Dispersions}
\label{sec:cvd}

\subsection{Isotropy}

The simplest assumption to make is that of isotropy ($\beta=0$).
Then $\sigma_r^2(r)$ is recovered uniquely from $\sigmalos^2(R)$
by an inverse Abel Transform. The Jeans equation reduces basically to
hydrostatic equilibrium with the ``pressure'' being equal to
$P=\rhoL\sigma^2$, that is,
%
\begin{equation}
\label{eq:hse}
\bnabla(\rhoL\sigma^2)=-\rhoL{\bnabla}\psi
\end{equation}
where $\sigma$ is the one-dimensional velocity dispersion of the
tracer population and $\psi$ is the gravitational potential.  

If we further assume that $\sigmalos^2(R)$ is a constant $\sigma_0^2$,
as suggested by most of the observations, then the unique solution is
$\sigma_r^2(r)=\sigma_0^2$. Under this assumption, the central
properties of the halo are severely restricted, as we now show.  Since
$\bnabla(\rhoL\sigma^2)=\sigma_0^2\bnabla\rhoL$, the Jeans equation
indicates that $\bnabla\rhoL$ and $\bnabla\psi$ are (anti-)parallel
everywhere. This further implies that the surfaces of constant $\rhoL$
and $\psi$ coincide, and thus $\rhoL$ can be considered as a function
of $\psi$. Consequently, $\bnabla\rhoL=(\mathrm d\rhoL/\mathrm
d\psi)\bnabla\psi$ and equation (\ref{eq:hse}) reduces to
\begin{equation}
\label{eq:deq}
\sigma_0^2\frac{\mathrm d\rhoL}{\mathrm d\psi}+\rhoL=0.
\end{equation}
By solving this differential equation, we find that
\begin{equation}
\label{eq:sol}
\rhoL=\rho_0\,\exp\!\left\lgroup{-\frac\psi{\sigma_0^2}}\right\rgroup
\,;\qquad
\psi=\psi_0-\sigma_0^2\ln\rhoL,
\end{equation}
where $\psi_0=\sigma_0^2\ln\rho_0$ is an integration constant.
Combined with the Poisson equation under the assumption that the
potential is generated by the dark matter halo of a density
$\rhoDM$ (i.e., $\rhoDM\gg\rhoL$ and so $\rho_{\rm t}\approx\rhoDM$),
we obtain
\begin{equation}
\label{eq:integdea}
\rhoDM=\frac{\lap\psi}{4\pi G}=
-\frac{\sigma_0^2}{4\pi G}\lap\ln\rhoL.
\end{equation}
This is an interesting equation, as the dark matter density, which we
wish to know, depends on the Laplacian of the luminosity density. The
combined integro-differential equations~(\ref{eq:integdeb}) and
(\ref{eq:integdea}) now relate the dark matter density to the
observables, $\sigma_0$ and $\SigmaL(R)$.

The implications of equations (\ref{eq:sol}) and (\ref{eq:integdea})
on the behaviour of tracer populations in dark matter halos are of
considerable interest, even before we apply to any specific
example. First, equation (\ref{eq:sol}) indicates that the potential
is finite for any finite luminosity density. Consequently, we find
that any cored luminosity density implies that the central potential
well cannot be infinitely deep and so the dark matter halo density
profile is also cored, or diverges strictly slower than the singular
isothermal sphere ($r^{-2}$) if it is cusped. Similar argument also
leads us to the conclusion that any cusped luminosity density would
only be supported by a cusped halo diverging at least as fast as a
singular isothermal sphere. In fact, this latter conclusion can be
sharpened with further analysis. Under the assumption of spherical
symmetry, equation (\ref{eq:integdea}) may be written to be
\begin{equation}
\label{eq:slope}
\gamma_\ell+
r\frac{\mathrm d\gamma_\ell}{\mathrm dr}
=\frac{4\pi G}{\sigma_0^2}\rhoDM r^2
\end{equation}
where $\gamma_\ell=-(\mathrm d\ln\rhoL/\mathrm d\ln r)$ is
the logarithmic slope of the density of the tracers. By taking
the limit to the centre ($r\rightarrow0$), we find that
\begin{equation}
\lim_{r\rightarrow0}\rhoDM r^2=
\frac{\sigma_0^2}{4\pi G}\gamma_{\ell,0},
\end{equation}
where $\gamma_{\ell,0}$ is the limiting value of $\gamma_\ell$ towards
$r\rightarrow0$, that is, the logarithmic cusp slope of the luminous
tracer density.  This indicates that if the luminosity density is
cusped with $\gamma_{\ell,0} >0$, the halo density must be cusped
as $\rhoDM\sim r^{-2}$ like a singular isothermal sphere. On the other
hand, any cored $\rhoL$ indicates that
$\gamma_{\ell,0}=\lim_{r\rightarrow0}(\rhoDM r^2)=0$ and thus the halo
density may not diverge as fast as or faster than the cusp of the
singular isothermal sphere. Finally, $\rhoL$ with a central hole
implies that $\gamma_{\ell,0}<0$ and is therefore unphysical as
it will lead to $\rhoDM<0$ \citep[c.f.,][]{AE06}


\subsection{Anisotropy}

In reality, the velocity dispersion tensor of a ``collisionless''
stellar system is not necessarily isotropic. However, provided the
gravitational potential is still spherically symmetric, the Jeans
equation reduces to
\begin{equation}
\label{eq:ssj}
\frac1I\frac{\mathrm d}{\mathrm dr}\!\left(I\rhoL\sigma_r^2\right)=
-\rhoL\frac{\mathrm d\psi}{\mathrm dr}
\end{equation}
where $I=\exp\int(2\beta/r)\mathrm dr$ is the integrating factor (e.g.,
$I=r^{2\beta}$ if $\beta$ is constant).

Inspired by the discussion in the preceding section, we now consider
the case of constant radial velocity dispersions with an arbitrary
functional form of $\beta$.  Note that it is possible, from equation
(\ref{eq:abelbeta}), to find $\beta(r)$ that is consistent with the
observed $\SigmaL(R)$ and $\sigmalos^2(R)$ once we ascribe a
particular behaviour to $\sigma_r^2(r)$. Hence, we can in principle
find such a model that produces a constant line-of-sight velocity
dispersion (or any other desired form), which is indicated by the
observations using equation (\ref{eq:abelbeta}).

If $\sigma_r^2=\sigma_0^2$ is a constant, equation (\ref{eq:ssj})
becomes
\[
\frac{\mathrm d\psi}{\mathrm dr}=
-\sigma_0^2\frac{\mathrm d}{\mathrm dr}\ln(I\rhoL)
\]
and consequently we find that
\begin{equation}
\psi=\psi_0-\sigma_{0}^2\ln(I\rhoL)
=\psi_0-\sigma_{0}^2\left(\ln\rhoL+\int\frac{2\beta}r\,\mathrm dr\right)
\label{eq:newpot}
\end{equation}
\begin{equation}
\rhoDM=\frac{\lap\psi}{4\pi G}
=-\frac{\sigma_0^2}{4\pi Gr^2}\,\frac{\mathrm d}{\mathrm dr}\!
\left\lgroup{r^2\frac{\mathrm d}{\mathrm dr}\ln(I\rhoL)}\right\rgroup.
\label{eq:newdens}
\end{equation}
Here, the last equation can be re-cast similarly to equation
(\ref{eq:slope}) so that
\begin{equation}
\gamma_\ell-2\beta
+r\left(\frac{\mathrm d\gamma_\ell}{\mathrm dr}
-2\frac{\mathrm d\beta}{\mathrm dr}\right)
=\frac{4\pi G}{\sigma_0^2}\rhoDM r^2.
\end{equation}
This is basically the same as equation (\ref{eq:slope}), except
$\gamma_\ell$ is replaced by $\gamma_\ell-2\beta$.  We
note that the result does not require the assumption that $\beta$ is a
constant. The implication of this result is quite similar to that of
equation (\ref{eq:slope}). At the limit towards the centre
($r\rightarrow0$), we find that $\gamma_{\ell,0}>2\beta_0$
implies that $\rhoDM\sim r^{-2}$ whilst we infer that
$\lim_{r\rightarrow0}(\rhoDM r^2)=0$ if $\gamma_{\ell,0}
=2\beta_0$.  Here, $\beta_0$ is the limiting value of
anisotropy parameter at the centre.  As per 
\citet{AE06}, $\gamma_{\ell,0}<2\beta_0$ is unphysical (although
$\rho_{\ell,0}$ does not self-consistently generate $\psi$, the
potential well depth is finite at the centre so that their result
holds) since it indicates negative halo density.

In summary, therefore, \emph{given a tracer population with a constant
  radial velocity dispersion, the generic solution of the Jeans
  equation for the dark matter is cusped like a singular isothermal
  sphere ($\rhoDM \propto r^{-2}$).  Milder cusps (like $\rhoDM
  \propto r^{-1}$) and cores are possible, but they are not
  generic. Such solutions only exist if the anisotropy parameter
  $\beta$ and the logarithmic slope of the stellar density
  $\gamma_\ell$ satisfy 
  $\gamma_{\ell,0}=2\beta_0$.}

\section{Examples}
\label{sec:ex}

\subsection{A Plummer Light Profile}

Plummer's law is commonly used to model the light of dSphs
\citep[e.g.,][]{La90,Wi02}. Assuming a constant mass-to-light ratio
$\Upsilon$, then the surface density is 
\begin{equation}
\SigmaL(R) = \frac{\Sigma_0}{(1+ R^2/r_0^2)^2}.
\end{equation}
Here, $r_0$ is the radius of the cylinder that encloses half the
light, whilst the total luminosity is $L = \pi r_0^2 \Sigma_0
/\Upsilon$.  It is straightforward to establish via
equation~(\ref{eq:integdeb}) that the stellar density is
\begin{equation}
\rhoL(r) = \frac{3 \Sigma_0}{4 r_0} \frac{1}{ (1+ r^2/r_0^2)^{5/2}}
\end{equation}
Now, using equation (\ref{eq:integdea}), the dark matter density must be
\begin{equation}
\rhoDM(r)  = \frac{5\sigma_0^2 }{ 4 \pi G r_0^2}
\frac{3 + r^2/r_0^2 }{ (1+ r^2/r_0^2)^2},
\end{equation}
which is a cored isothermal sphere \citep[see][]{Ev93}.  
However, as the model is isotropic ($\beta
=0$) and the stellar density is cored at the
centre ($\gamma_{\ell,0}=0$), this dark halo solution corresponds to
the special case $\gamma_{\ell,0}=2\beta_0$.

Now suppose the assumption of isotropy is dropped. Using
equation~(\ref{eq:newdens}), we find that the dark halo density
acquires an additional term that behaves near the centre as
\begin{equation}
\label{eq:densextra}
  \rhoDM(r)\simeq-{\beta_0 \sigma_0^2 \over 2 \pi G r^2},
\end{equation}
from which we deduce that $\beta_0 \le0$ so that the model is
tangentially anisotropic as $r \rightarrow 0$.  Notice that this has
changed the behaviour of the density at the origin -- the halo law has
now an isothermal cusp. This is in accord with the general result that
provided $\gamma_{\ell,0}>\beta_0$, the cusp is isothermal.

\subsection{Exponential Light Profiles}

Another set of profiles often used to model dSph light distributions
is based on the exponential law \citep{Se68,Fa83,Ir95}.
Suppose that the three-dimensional density law $\rhoL(r)$
is exponential
%
\begin{equation}
\rhoL(r) = \rho_0\, \exp\!\left\lgroup{-\frac r{\rd}}\right\rgroup.
\end{equation}
The corresponding surface density is cored as
\begin{equation}
\SigmaL(R) = \Sigma_0\ \frac R{\rd}\
K_1\!\left\lgroup{\frac R{\rd}}\right\rgroup
\label{eq:bessel}
\end{equation}
where $\Sigma_0 = 2\rd\rho_0$ and $K_\nu(x)$ is the modified Bessel
function of the second kind and of order $\nu$. We refer to this as
the Bessel profile.  The half-light radius is $\sim2.027\rd$, whilst
the total luminosity is $L = 4\pi \rd^2 \Sigma_0 /\Upsilon$.  The dark
matter density inferred from equation (\ref{eq:integdea}) is then
\begin{equation}
\rhoDM = \frac{\sigma_0^2}{2 \pi G \rd r},
\end{equation}
which is the $r^{-1}$ cusp beloved of cosmologists \citep[e.g.,][]{NFW}.
In fact, a three-dimensional density law $\rhoL=\rho_0\mathrm
e^{-(r/\rd)^\alpha}$ leads to an infinite dark matter cusp of form
$\rhoDM\propto r^{-(2-\alpha)}$.

Suppose instead the surface brightness profile is modelled with an
exponential law
%
\begin{equation}
\SigmaL(R) = \Sigma_0\, \exp\!\left\lgroup{ -\frac R{\Rd} }\right\rgroup.
\end{equation}
The luminosity density is then \citep[see e.g.,][]{Ke92}
%
\begin{equation}
\rhoL(r) = \frac{\Sigma_0}{\pi\Rd}\
K_0\!\left\lgroup {\frac r{\Rd} }\right\rgroup,
\end{equation}
which is logarithmically divergent at the centre.
The dark matter density is now
%
\begin{equation}
\rhoDM(r)=\frac{\sigma_0^2}{4\pi G\Rd r}
\left[\left(\frac{[K_1]^2}{[K_0]^2}-1\right)\frac r\Rd+\frac{K_1}{K_0}\right]
\end{equation}
where $K_n=K_n(r/\Rd)$ and $n=0,1$.
It is singular as $r \rightarrow 0$, 
%
\begin{equation}
\rhoDM \simeq \frac{\sigma_0^2}{4\pi G} \frac{1}{r^2\ln(r^{-1})},
\end{equation}
which exhibits strictly slower divergence than a singular isothermal
sphere.  In other words, once the luminosity density has been assumed
to be of exponential form (either in projection or in
three-dimensions), then the inferred dark matter density is
cusped, but the cusp is always weaker than the isothermal cusp
$\rhoDM \sim r^{-2}$.

It is easy to see that equation~(\ref{eq:slope}) still applies since
$\gamma_\ell\rightarrow0$ as $r\rightarrow0$.  Again, the isotropic
models are somewhat unusual -- the introduction of anisotropy drives
the dark halo solution towards an isothermal cusp.  The fact that the
terms in the density and the anisotropy decouple in
equation~(\ref{eq:newdens}) means that the assumption of any central
anisotropy gives the same additional contribution to the halo
density~(\ref{eq:densextra}) as in the previous example.

\section{The General Case}
\label{sec:gen}

Let us now gain insight into the general case by discarding the
assumption that the radial velocity dispersion is constant. We derive
the extension of our result to the general spherically symmetric case.
By rewriting the spherical steady-state Jeans equation~(\ref{eq:ssj0}),
we obtain (under the assumption that $\rho_{\rm t}=\rhoDM$)
%
\begin{equation}
4\pi G\int_0^r\!\rhoDM(\tilde r)\,\tilde r^2\,\mathrm d\tilde r
=r \sigma_r^2 \left( \gamma_\ell - 2 \beta
 - \frac{\mathrm d \ln \sigma_r^2}{\mathrm d \ln r}\right),
\end{equation}
and 
\begin{equation}
4\pi G \rhoDM r^2 =
\sigma_r^2 \left( 1 + \frac{\mathrm d \ln \sigma_r^2}{\mathrm d \ln r}
+ r \frac{\mathrm d}{\mathrm dr}\right)
 \left( \gamma_\ell - 2\beta
 - \frac{\mathrm d \ln \sigma_r^2}{\mathrm d \ln r} \right).
\label{eq:generalcase}
\end{equation}

Now note that if the dark matter density has an isothermal or steeper
cusp, then the left-hand side of equation~(\ref{eq:generalcase}) tends
to a non-zero value as $r \rightarrow 0$. 
However, if the dark matter density is cored or diverges more slowly
than $r^{-2}$, then the left-hand side vanishes as $r \rightarrow 0$.
Then, for the right-hand side also to vanish as $r \rightarrow 0$, one
of the following three conditions must hold at the same limit
\begin{enumerate}
\item\[
\frac{\mathrm d \ln \sigma^2_r}{\mathrm d \ln r}
\rightarrow -1
\]\smallskip
\item\[
\frac{\mathrm d \ln \sigma^2_r}{\mathrm d \ln r}
\rightarrow\gamma_\ell - 2 \beta
\]\smallskip
\item $\sigma_r^2 \rightarrow 0$.
\end{enumerate}
Case (i) implies that $\sigma_r^2 \sim r^{-1}$. Excluding the
possibility that there is a black hole at the centre, then the central
potential must be finite or divergent strictly slower than the
logarithm (as $\rhoDM r^2 \rightarrow 0$), and the velocity dispersion
diverging as a power law cannot be supported. Case (ii) implies that
$\sigma_r^2 \sim r^{{\gamma_\ell}- 2 \beta}$. If $\gamma_\ell >
2\beta$, then $\sigma_r^2 \rightarrow 0$ and so this may be subsumed
into Case (iii). If $\gamma_\ell < 2 \beta$, then $\sigma_r^2$
diverges as a power law and so is again unphysical. This leaves only
$\gamma_\ell = 2 \beta$ as an independent possibility.

Consequently, we have established that if the dark matter density is
cored or falls off with a cusp less severe than $r^{-2}$, then either
$\gamma_\ell = 2 \beta$ or $\sigma_r^2 =0$ at the centre. The
discarding of the assumption of constancy of the radial velocity
dispersion permits one additional possibility, namely that
$\sigma_r^2$ falls to zero at $r=0$.

In summary, therefore, we have a surprisingly strong and general
result. \emph{Given a tracer population in the spherical Jeans
  equation, the generic solution for the dark matter is cusped
  like a singular isothermal sphere ($\rhoDM \propto r^{-2}$).  Milder
  cusps and cores are possible, but they are not generic. Such
  solutions exist either if the anisotropy parameter $\beta$ and the
  logarithmic slope of the stellar density $\gamma_\ell$ satisfy
  $\gamma_{\ell,0}=2\beta_0$ or if the central radial velocity
  dispersion $\sigma^2_{r,0}$ vanishes.}

Such a strong result may seem astonishing.  It is worth remarking that
this result is essentially due to the spherical symmetry assumption,
and in particular, the coordinate singularity at the origin. A real
dSph is not likely to be exactly spherically symmetric at the
centre. So, it may be possible for a dSph to have a finite velocity
dispersion towards the centre by mild symmetry-breaking. Outside of
the very central region, a dSph can often be well-approximated by the
idealized spherically symmetric fiction.  Our theorem therfore really
lays bare the dangers of inferring the central density profile from
the almost universally made assumption of spherical symmetry.

\section{Summary}

We have studied the problem of deducing the dark halo density from the
surface brightness and velocity dispersion profiles of a tracer
population. This has immediate application to the dwarf spheroidals
satellite galaxies (dSphs) of the Milky Way.

If the stellar population generates the gravity field, then the Jeans
and Poisson equations give a unique solution for the density, the
mass-to-light ratio and the anisotropy of the spherical model,
consistent with the observed surface brightness and the line-of-sight
velocity dispersion \citep{Bi82,To83}. Of course, this does not apply to
the case of dSphs, in which the density of the stellar population is
dominated by the dark halo density. Now, the problem suffers from the
well-known mass-anisotropy degeneracy.

The line-of-sight velocity dispersion profiles of the Milky Way dSphs
appear to be usually flat \citep[see e.g.,][]{Kl01,Kl02,Ko07a,Wa07},
which suggests the simple assumption that the velocity dispersion
tensor is isotropic and has a constant value. Then, any inference as
to the central behaviour of the dark matter potential is controlled by
the assumption as to the light profile of the tracer population.  If
the light profile is cored, then a dark matter halo density that is
itself cored is deduced from Jeans modelling. If the light profile is
cusped like an exponential law (or its variants), then a dark halo
density that has a milder cusp than isothermal (such as the
Navarro-Frenk-White cusp of $\rhoDM \propto r^{-1}$) is deduced.  This
provides the explanation as to why previous investigators
\citep{Ko07b, Wi06} have concluded that the data are consistent with
both cusps and cores.

However, velocity anisotropy in the stellar population has a dramatic
effect on the dark halo density recovered from Jeans modelling.  If a
tracer population has a constant radial velocity dispersion, then the
generic solution for the dark halo is always cusped like a singular
isothermal sphere ($\rhoDM \propto r^{-2}$).  Milder dark matter cusps
(like $\rhoDM \propto r^{-1}$) and cores are possible, but they are
not generic.  They can occur only when the condition $\gamma_\ell =
2\beta$ is fulfilled at the centre, where $\beta$ is the anisotropy
parameter and $\gamma_\ell$ is the logarithmic slope of the stellar
density.  Note that many of the commonly used dSph models (such as
Plummer or exponential profiles with isotropic velocities) correspond
to the special case $\gamma_\ell = 2\beta$ and so any conclusions
inferred as to the dark halo law may not be beyond reproach.

Finally, even if the assumption as to the constancy of the radial
velocity dispersion is discarded, then almost the same theorem holds
true. If $\gamma_\ell = 2\beta$ at the centre or if $\sigma_r^2$ falls
to zero at the centre, then dark matter cores and milder cusps than
isothermal are possible. The generic solution, however, remains the
isothermal dark matter cusp, at least within the framework of
the spherical symmetric Jeans and Poisson equations.

Here, our examples and analysis have shown how Jeans solutions may be
telling modellers more about their assumptions rather than the
theoretical implications of the data. Of course, there is in
principle more information in the discrete velocities than in the
velocity dispersion profile and the Jeans equations. The best response
to the degeneracy of the problem is to seek further observational
constraints (perhaps higher moments from the line profile, see e.g.,
\citealt{MK90}) or additional insights on the behaviour of the
anisotropy (maybe from simulations or from a detailed analysis on the
underlying physics).

\section*{acknowledgments}
JA acknowledges that the Dark Cosmology Centre is funded by the Danish
National Research Foundation (Danmarks Grundforskningsfond). We thank
an anonymous referee for a number of stimulating questions.

\label{lastpage}
\end{document}